# Bulk topological insulators as inborn spintronics detectors

Y. Shiomi[1], K. Nomura[1], Y. Kajiwara[1], K. Eto[2], M. Novak[2], Kouji Segawa[2], Yoichi Ando[2], and E. Saitoh[1,3,4,5]

1. Institute for Materials Research, Tohoku University, Sendai 980-8577, Japan
2. Institute of Scientific and Industrial Research, Osaka University, Ibaraki, Osaka 567-0047, Japan
3. WPI Advanced Institute for Materials Research, Tohoku University, Sendai 980-8577, Japan
4. CREST, Japan Science and Technology Agency, Tokyo 102-0076, Japan
5. Advanced Science Research Center, Japan Atomic Energy Agency, Tokai 319-1195, Japan

**Detection and manipulation of electrons' spins are key prerequisites for spin-based electronics or spintronics [1]. This is usually achieved by contacting ferromagnets with metals or semiconductors, in which the relaxation of spins due to spin-orbit coupling limits both the efficiency and the length scale. In topological insulator materials [2], on the contrary, the spin-orbit coupling is so strong that the spin direction uniquely determines the current direction, which allows us to conceive a whole new scheme for spin detection and manipulation [3,4]. Nevertheless, even the most basic process, the spin injection into a topological insulator from a ferromagnet, has not yet been demonstrated. Here we report successful spin injection into the surface states of topological insulators by using a spin pumping technique. By measuring the voltage that shows up across the samples as a result of spin pumping, we demonstrate that a spin-electricity conversion effect takes place in the surface states of bulk-insulating topological insulators $Bi_{1.5}Sb_{0.5}Te_{1.7}Se_{1.3}$ [5,6] and Sn-doped $Bi_2Te_2Se$ [7]. In this process, due to the two-dimensional nature of the surface state, there is no spin current along the perpendicular direction. Hence, the mechanism of this phenomenon is different from the inverse spin Hall effect [8] and even predicts perfect conversion between spin and electricity at room temperature. The present results reveal a great advantage of topological insulators as inborn spintronics devices.**

The spin pumping technique has proved versatile [8-16] for injecting spins into materials as a result of magnetization *M* precession [see Supplementary Information

(SI), Sec. A]. In a bilayer film consisting of a ferromagnet on top of a paramagnet, magnetization precession in the ferromagnet generates non-equilibrium spins in the paramagnet near the interface upon excitations of ferromagnetic resonance (FMR) [10]. The spin pumping also exerts torque on the precessing magnetization in the ferromagnet and enhances the damping constant for magnetization precession (Gilbert damping constant) [8-17]. The spin polarization in the paramagnet decays due to spin relaxation, the length scale of which is characterized by the spin diffusion length $\lambda$ [18].

Topological insulators (TIs) are a new class of quantum materials that possess topologically-protected metallic surface states [2-4], while the interior is insulating (Fig. 1**a**). Furthermore, conduction electrons on the surface states behave as Dirac fermions that bear a special characteristic called *spin-momentum locking* [2-4]; namely, in the surface states of TIs, the direction of the electron's motion uniquely determines its spin direction and vice versa. Examples are shown in Fig. 1**b**: here, at the Fermi level, right-moving (+*x* direction) and left-moving (-*x* direction) electrons have spins pointing to +*y* and -*y* (or -*y* and +*y*) directions, respectively. Hence, if a spin imbalance is induced in the surface state by spin pumping, a charge current $\boldsymbol{J}_c$ is expected to show up along the "Hall" direction defined by

$$\boldsymbol{J}_c \parallel (\boldsymbol{z} \times \boldsymbol{\sigma}), \tag{1}$$

where $\boldsymbol{\sigma}$ is the direction of the spin polarization and $\boldsymbol{z}$ is the unit vector perpendicular to the plane. Since the spin-momentum locking is the origin of this spin-electricity conversion effect, 100% conversion is possible in principle. Here, the spin polarization is induced not as a result of spin current but through the direct exchange coupling with the ferromagnet (see SI, Sec. A), and hence the mechanism of this spin-electricity conversion is different from that in the inverse spin Hall effect.

$Bi_{1.5}Sb_{0.5}Te_{1.7}Se_{1.3}$ (BSTS) is a bulk TI in which electric conduction through its surface state is dominant at low temperature, thanks to the nearly perfect carrier compensation in the bulk part [5,6,19,20]; In contrast, in stoichiometric compounds like $Bi_2Se_3$ and $Bi_2Te_3$, a sizable amount of bulk carriers due to unintentional doping from crystalline defects are inevitable [3,6,21]. The details of the topological surface states in BSTS have been elucidated by angle-resolved photoemission spectroscopy [22]. We also studied Sn-doped $Bi_2Te_2Se$ (Sn-BTS), which was recently reported to be even more bulk-insulating than BSTS [7]. Currently, BSTS and Sn-BTS are the best bulk TI materials for studies of the topological surface states [3].

Figure 1**c** shows a schematic illustration of the present experiment. A 20-nm-thick $Ni_{81}Fe_{19}$ (permalloy, Py) film was deposited in a high vacuum on the middle part of a cleavage plane of a bulk TI crystal (thickness ~0.1 mm). Microwave was applied to the sample through a vector network analyzer in a static magnetic field. A dc electromotive force was measured on the TI sample along the direction perpendicular to the applied magnetic field, while the frequency was kept at 5 GHz.

Figure 1**d** shows the temperature ($T$) dependence of the resistivity ($\rho$) for the samples used in the present study. We measured three BSTS samples (BSTS1, BSTS2, and BSTS3) and one Sn-BTS sample, as well as two $Bi_2Se_3$ (BS) samples (BS1 and BS2) for comparison. The resistivity of all the BSTS and Sn-BTS samples monotonically increases with decreasing $T$ below 200 K, which indicates that the bulk conduction is well suppressed at low temperature [5-7,19,20]. In contrast, the resistivity of the BS samples shows a metallic $T$-dependence with a small magnitude of ~$10^{-4}$ $\Omega$cm; the bulk conduction is dominant in the BS samples [6].

Figure 2**a** shows the magnetic field ($H$) dependence of the amplitude of the microwave transmittance through the sample, $|S_{21}|$, which presents negative peaks around $\mu_0 H = \pm 30$ mT corresponding to the FMR of $Ni_{81}Fe_{19}$. At the FMR condition of the $Ni_{81}Fe_{19}$ layer ($H \equiv \pm H_r$), the absorption of microwave in $Ni_{81}Fe_{19}$ causes a decrease in $|S_{21}|$. The line width of the FMR spectrum carries important information on the relaxation of magnetization, or the Gilbert damping constant ($\alpha$) [23]. In Fig. 2**b**, we show the frequency ($f$) dependence of the half-maximum full-width (HMFW) of the FMR spectrum ($\Delta H_i \propto f\alpha$) for BSTS|$Ni_{81}Fe_{19}$ and for $Ni_{81}Fe_{19}$ alone at 15 K (see also Supplementary Fig. **S1**). As shown in Fig. 2**b**, the linear $f$-dependence of $\Delta H_i$ is observed for our samples and the slope ($\propto \alpha$) for BSTS|$Ni_{81}Fe_{19}$ is greater than that for $Ni_{81}Fe_{19}$ alone. The increase in $\alpha$ in the presence of BSTS gives evidence for spin injection into the attached BSTS from $Ni_{81}Fe_{19}$ [8-17]; to the best of our knowledge, this is the first experimental realization of spin injection from a ferromagnet into a TI.

In Fig. 2**c**, we show the results of the electromotive-force measurement with external magnetic field ($H$) at various temperatures for BSTS1|$Ni_{81}Fe_{19}$. At the FMR condition ($H = \pm H_r$), a sharp peak (or dip) in the voltage ($V$) arises (Fig. 2**c**). At 75 K, the peak height $V_0$ (measured from the background) is almost symmetric between -$H_r$ and +$H_r$, and its absolute magnitude reaches ~0.5 mV, while it decreases with decreasing $T$. Below 39 K, the voltage signal presents an asymmetric shape and $V_0$ is different for –$H_r$ and +$H_r$. This asymmetry becomes more pronounced at lower temperature, and below 27 K, even a sign reversal between -$H_r$ and +$H_r$ is observed. This sign reversal is what is expected from eq. (1), since the direction of the injected spin reverses between -$H_r$ and +$H_r$.

We note that this signal is irrelevant to the voltage generation inside the $Ni_{81}Fe_{19}$ film, since such a Hall-type signal was not observed in control samples. In Fig, 3**a**, we show voltage signal near the FMR of $Ni_{81}Fe_{19}$ for BSTS1, BSTS2, BS1, and *n*-type Si of the same size. Here, as an example of conventional conductors, we used *n*-type Si (*n*-Si) whose $\rho$ is ~1 $\Omega$cm with $d\rho/dT$~0.01 $\Omega$cm/K (>0) at 295 K; the magnitude of $\rho$ (~1 $\Omega$cm) is almost the same as that for BSTS at 15 K (Fig. 1**d**). Though a slight increase in $\alpha$ indicates spin injection into *n*-Si (Fig. 2**b**), $V_0$ at FMR for *n*-Si|$Ni_{81}Fe_{19}$ is clearly symmetric between $-H_r$ and $+H_r$; no antisymmetric signal is observed. Also for BS1, where its bulk conduction is much more prominent than the surface conduction, antisymmetric signal is absent. The symmetric signal, whose sign and magnitude seem to be random among BS or *n*-Si samples, can be attributed to the bulk Seebeck effect due to a small temperature-gradient along the sample plane that is peaked at FMR. By contrast, not only in BSTS1|$Ni_{81}Fe_{19}$ but also in BSTS2|$Ni_{81}Fe_{19}$, clear antisymmetric signals are observed. These results indicate that the antisymmetric signal observed in the BSTS samples is generated in the surface state of BSTS.

The microwave-power ($P_{in}$) dependence of the voltage signal also confirms that the electromotive force for BSTS is associated with the surface state. Figure 3**b** shows the $H$ dependence of $V$ at several $P_{in}$ values for BSTS2|$Ni_{81}Fe_{19}$. The sign reversal is observed at all $P_{in}$ values and the magnitude increases with increasing microwave power ($P_{in}$). The values of $V_0$ at 15 K and 39 K are shown in Fig. 3**c** as a function of microwave absorption power at FMR, $\Delta P$ (see also Supplementary Fig. **S2**). At 39 K, $V_0$ shows a nonlinear dependence on the absorption power ($\Delta P$); this is attributed to the Seebeck effect in the bulk state of BSTS, because the $\Delta P$ dependence of the Seebeck voltage ($=S\Delta T$) is expected to be nonlinear due to the variation of the Seebeck coefficient, $S$, by the heating effect caused by FMR. Since $S(T) \propto T$ (Mott relation [24]) and $\Delta T \propto \Delta P$, one obtains $S\Delta T \propto (\Delta P)^2$. The antisymmetric (odd) signal observed at 15 K is, by contrast, proportional to $\Delta P$ (i.e. $V \propto \Delta P$, see also Supplementary Fig. **S3**), which clearly indicates that the antisymmetric signal is of different origin.

The magnitude of $V_0$ is almost the same between $-H_r$ and $+H_r$ for BSTS2|$Ni_{81}Fe_{19}$, while it is somewhat different in BSTS1|$Ni_{81}Fe_{19}$ (Fig. 3**a**). We separate the antisymmetric part from the symmetric part in $\tilde{V} = V_0 / \Delta P$ at 15 K by calculating $\tilde{V}^a \equiv [\tilde{V}(H) - \tilde{V}(-H)]/2$. Figure 4**a** shows resulting $\tilde{V}^a$ for the three BSTS|$Ni_{81}Fe_{19}$ samples, as well as for the BS|$Ni_{81}Fe_{19}$ control samples, as functions of $H$. The sign of $\tilde{V}^a$ is positive at $-H_r$ and negative at $+H_r$ for all the BSTS|$Ni_{81}Fe_{19}$ samples. The magnitude of $\tilde{V}^a$ is similar among the three BSTS|$Ni_{81}Fe_{19}$ samples, but it tends to be larger in samples with larger $\rho$ (see Fig. 1**d**). In the BS samples (BS1|$Ni_{81}Fe_{19}$ and BS2|$Ni_{81}Fe_{19}$), the symmetric part is dominant as shown in Fig. 3**a**, and $\tilde{V}^a$ is negligibly small.

We have formulated a theoretical model for the spin-electricity conversion effect on the topological surface state, as illustrated in Fig. 4**b** (see SI, Sec. B, for details). The spin polarization on the surface state, <$\sigma_y$>, is induced by the spin pumping driven by FMR in $Ni_{81}Fe_{19}$ [10,16,17,25]. Owing to the spin-momentum locking in the topological surface state, the spin polarization per unit area, <$\sigma_y$>/$A$ ($A$: area), gives rise to a shift of the Fermi circle of the helical Dirac fermions toward the $x$ direction, which induces the electric field along the "Hall" direction according to $E_x = -\frac{4\pi\hbar}{ek_F\tau}\frac{\langle\sigma_y\rangle}{A}$ (Fig. 4**b**) in the diffusive transport regime [26] (SI, Sec. B). Here, $k_F$ and $\tau$ are the Fermi momentum and the scattering time of the helical Dirac fermions, respectively. The sign of the produced voltage is consistent with the experimental results (Figs. 4**a** and 4**b**). The induced spin polarization is related to the spin pumped rate $\Sigma$ via $\langle\sigma_y\rangle = \eta\Sigma\tau$ (SI, Sec. B). Here, $\eta$ is a phenomenological parameter quantifying the spin-injection efficiency affected by spin flip and leakage into residual bulk carriers, by reflection of spins to the ferromagnet, and also by the effective spin-exchange coupling at the interface. From the experimental data, the value of $\eta$ is estimated to be $\sim 10^{-4}$ for BSTS1 and BSTS2.

Note that, since the spin pumping induces spin polarization directly on the Fermi level of the surface state, the voltage generation is obviously dominated by the surface spin polarization, in contrast to experiments with circularly polarized light [27,28] where interband excitations generates electron-hole pairs and photocurrents. Also, as already noted, the spin-electricity conversion effect in the present surface-electron system is different from the inverse spin Hall effect [8-16], in which the spin current ($J_s$) flowing in a paramagnet is converted into a charge current. It is worth mentioning that the present spin-electricity conversion effect is rather similar to that proposed in Rashba-split systems [29], but the efficiency is much higher in TIs than that in the Rashba-split system. The spin-momentum locking on a single Dirac cone predicts complete spin-electricity conversion in TIs even at room temperature as long as the surface state is robust, while in the Rashba-split systems where a pair of bands exist, one of the bands always counteracts the effect of the other.

Finally, to firm up our interpretation of the origin of the spin-electricity conversion effect, we show in Fig. 4**c** the data for a Sn-BTS sample possessing a large activation gap for the bulk conduction channel (Fig. 1**d**) [7]. In line with the large $\rho$ value that extends to higher temperatures than in BSTS samples, the sign reversal in $V_0$ between $-H_r$ and $+H_r$ is observed at high temperatures, even at 160 K. Figure 4**d** shows the $T$ dependences of symmetric ($V^s \equiv [V(H)+V(-H)]/2$) and antisymmetric ($V^a \equiv [V(H)-V(-H)]/2$) parts of $V$ at temperatures below 220 K, where $V$ is well fitted using a Lorentz function. The symmetric part $V^s$ changes sign at ~160 K, which should be attributed to a sign change in the bulk Seebeck coefficient if $V^s$ is due to the Seebeck effect in the bulk sate. Indeed, the Hall coefficient ($R_H$) changes sign around this temperature (Fig. 4**e**), below which the surface transport becomes dominant [7]. The clear correlation between the onset of surface dominated transport and the growth of the antisymmetric signal $V^a$ below ~160 K gives strong support to the interpretation that $V^a$ is a result of the spin-electricity conversion in the surface state.

Methods:

The single crystals of $Bi_{1.5}Sb_{0.5}Te_{1.7}Se_{1.3}$, $Bi_2Se_3$, and Sn-doped (0.4%) $Bi_2Te_2Se$ were grown by a Bridgman method in evacuated quartz tubes [5-7]. 20-nm-thick $Ni_{81}Fe_{19}$ thin films were deposited in a high vacuum by electron-beam evaporation on cleaved surfaces of TIs. The sizes of the BSTS and Sn-BTS samples were $4\times1\times0.1$ mm$^3$ (BSTS1), $4\times3\times0.1$ mm$^3$ (BSTS2), $2\times1.5\times0.2$ mm$^3$ (BSTS3), and $2.5\times1\times0.3$ mm$^3$ (Sn-BTS). We performed FMR and dc-voltage measurements for the TI|$Ni_{81}Fe_{19}$ and $n$-Si|$Ni_{81}Fe_{19}$ samples by using coplanar waveguides. The microwave was applied through the waveguides in a static magnetic field and the frequency was kept at 5 GHz in the measurement of the spin-electricity conversion effect. The in-plane resistivity and Hall resistivity were measured down to 2 K by the conventional four-probe method in a superconducting magnet.

Acknowledgement:

We thank G. Tatara, Y. Fujikawa, K. Ando, and R. Iguchi for fruitful discussions and T. An and Z. Qiu for experimental help. This work was supported by CREST-JST ("Creation of Nanosystems with Novel Functions through Process Integration"), JSPS (NEXT Program and KAKENHI No. 24244051, No. 24740211, No. 25220708, and for JSPS Fellows), AFOSR (AOARD 124038), and MEXT (Innovative Area "Topological Quantum Phenomena" (No. 22103004, No. 25103702, and No. 25103703)).

Author contributions:

Y.S. carried out the experiments with assistance from Y.K., and K.E., M.N., K.S., and Y.A. synthesized and characterized the samples. K.N. formulated the theoretical model on the experimental results. Y.S. analyzed the data. Y.A. and E.S. conceived the project. Y.A. supervised the materials part and E.S. supervised the measurement and analysis part. Y.S. and Y.A. wrote the manuscript.

Correspondence should be addressed to Y.S. (e-mail: shiomi@imr.tohoku.ac.jp) and Y.A. (e-mail: y_ando@sanken.osaka-u.ac.jp)

<u>Additional information:</u>

The authors declare no competing financial interests. Supplementary information accompanies this paper.

Figure 1

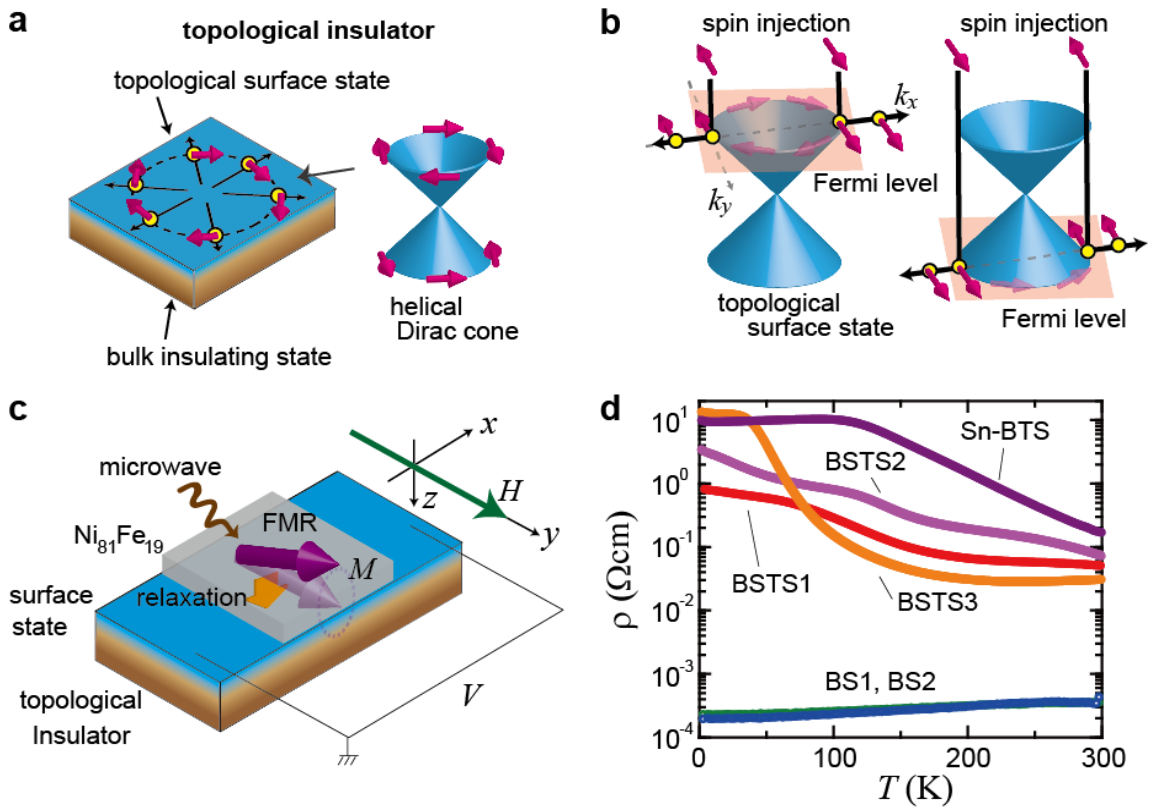

Figure 1: **Concept, Experimental setup, and samples**

(**a**) Schematic illustrations of a topological insulator, which has a conducting surface state consisting of helical Dirac fermions. The spin direction is determined uniquely by the direction of electron motion (spin-momentum locking). (**b**) Concept of spin-electricity conversion effects on the topological surface state. When spin is injected, the spin-momentum locking results in a charge current along the "Hall" direction on the surface state. (**c**) A schematic illustration of the experiment of the spin-electricity conversion effects. (**d**) Temperature ($T$) dependence of the resistivity ($\rho$) for $Bi_2Se_3$ (BS), $Bi_{1.5}Sb_{0.5}Te_{1.7}Se_{1.3}$ (BSTS), and Sn-doped $Bi_2Te_2Se$ (Sn-BTS) samples.

Figure 2

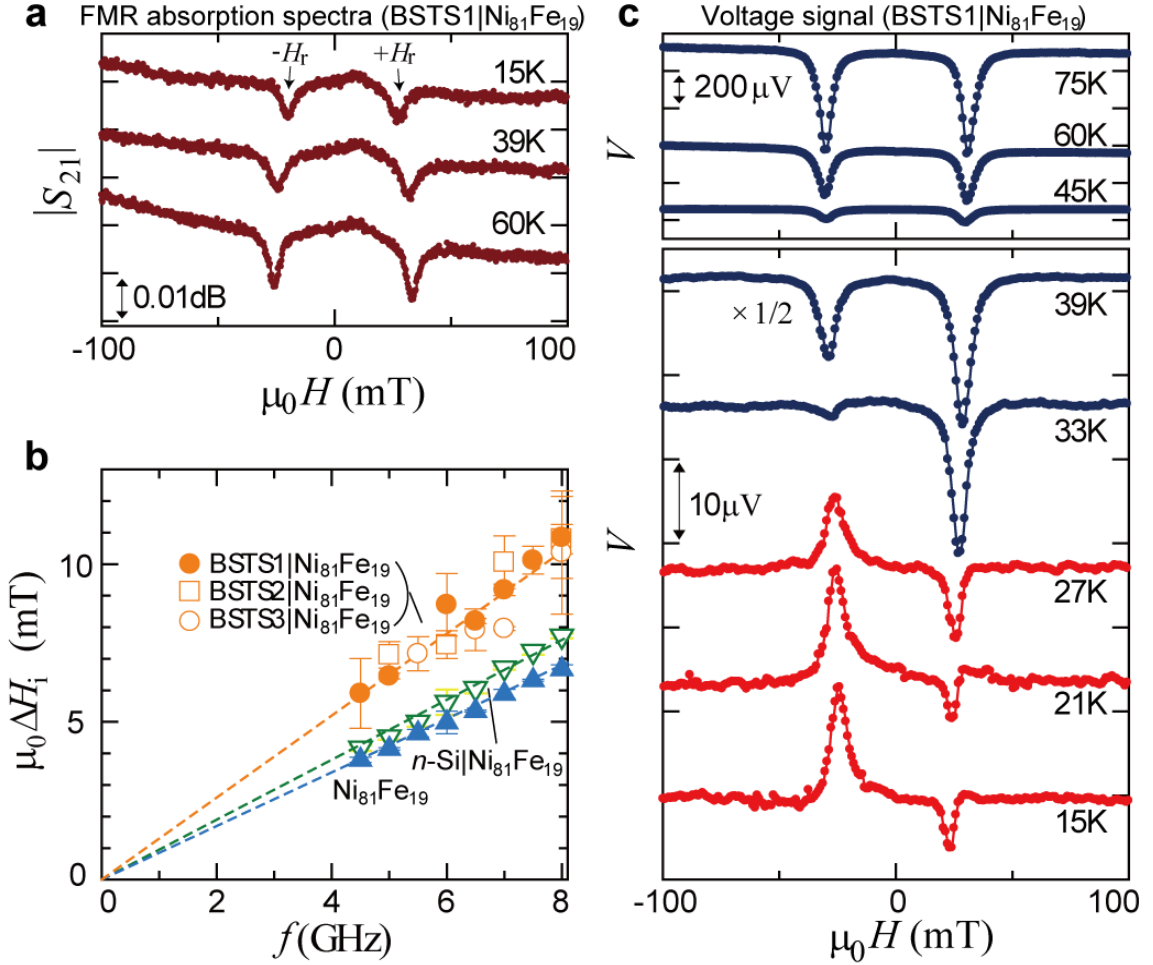

Figure 2: **Temperature variation of electromotive force for BSTS1|Ni$_{81}$Fe$_{19}$**

(**a**) Magnetic-field ($H$) dependence of the FMR spectrum ($|S_{21}|$) at several temperatures for BSTS1|Ni$_{81}$Fe$_{19}$. At FMR, $|S_{21}|$ exhibits negative peaks due to the microwave absorption. (**b**) Frequency ($f$) dependence of the half-maximum full-width (HMFW, $\Delta H$) of the FMR spectrum for BSTS|Ni$_{81}$Fe$_{19}$ samples, $n$-Si|Ni$_{81}$Fe$_{19}$, and Ni$_{81}$Fe$_{19}$ film alone. $\Delta H_i$ (=$\Delta H-\Delta H_0$) is the intrinsic part of HMFW, where $\Delta H_0$ is the extrinsic part of HMFW originating from the sample inhomogeneity (see also Supplementary Information). The broken lines are merely guides to the eyes. (**c**) Magnetic-field ($H$) dependence of the electromotive force ($V$) at various temperatures for BSTS1|Ni$_{81}$Fe$_{19}$; the data are shifted for clarity. The microwave power ($P_{in}$) was 0.4 mW and the frequency ($f$) was 5 GHz. The antisymmetric (odd) signal appears below ~40 K, causing a sign reversal in $V_0$ between $-H_r$ and $+H_r$ below 27 K.

Figure 3

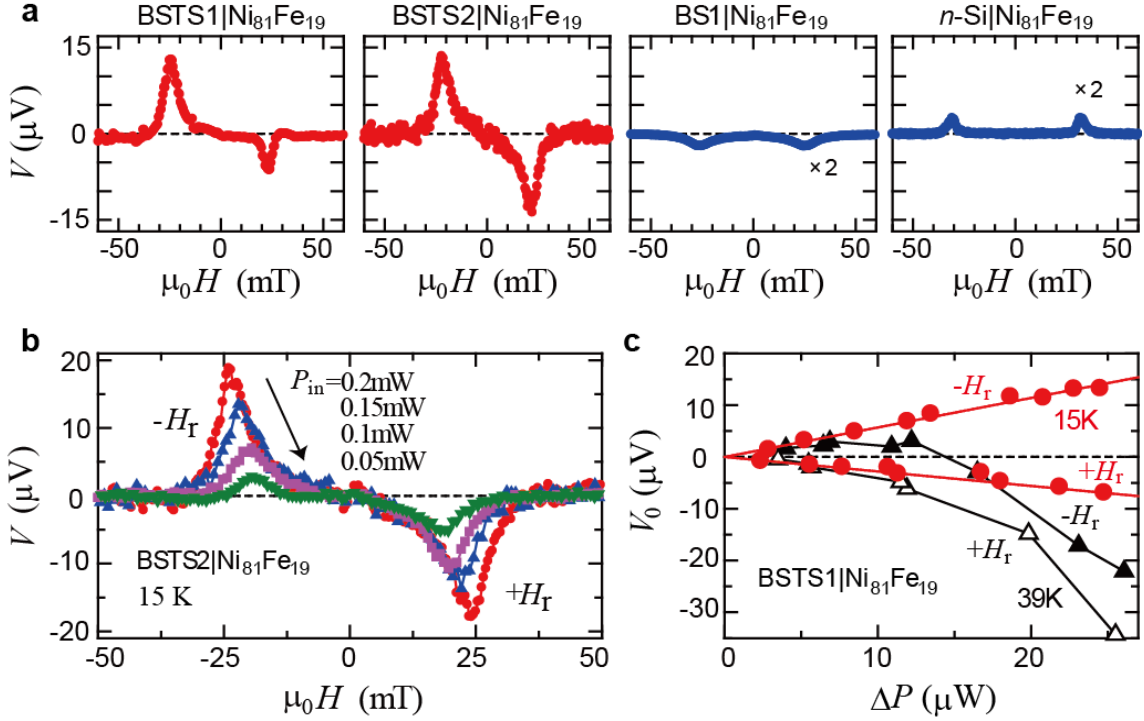

Figure 3: **Electromotive force for BSTS|Ni$_{81}$Fe$_{19}$ samples at 15 K**

(**a**) Electromotive force (*V*) for BSTS1|Ni$_{81}$Fe$_{19}$, BSTS2|Ni$_{81}$Fe$_{19}$, and BS1|Ni$_{81}$Fe$_{19}$ at 15 K and for *n*-Si|Ni$_{81}$Fe$_{19}$ at 295 K. The voltage signal is almost symmetric (even) between –*H*$_r$ and +*H*$_r$ for BS1|Ni$_{81}$Fe$_{19}$ and *n*-Si|Ni$_{81}$Fe$_{19}$, while a clear antisymmetric (odd) contribution is observed for both BSTS1|Ni$_{81}$Fe$_{19}$ and BSTS2|Ni$_{81}$Fe$_{19}$. The microwave power (*P*$_{in}$) was 0.4 mW (BSTS1|Ni$_{81}$Fe$_{19}$), 0.15 mW (BSTS2|Ni$_{81}$Fe$_{19}$), 0.4 mW (BS1|Ni$_{81}$Fe$_{19}$), and 0.13 mW (*n*-Si|Ni$_{81}$Fe$_{19}$). The frequency (*f*) was 5 GHz. (**b**) Magnetic-field (*H*) dependence of electromotive force (*V*) at 15 K for BSTS2|Ni$_{81}$Fe$_{19}$ at several values of microwave power (*P*$_{in}$). The frequency (*f*) was 5 GHz. (**c**) The absorption power (Δ*P*) dependence of *V*$_0$ (peak value of *V* measured from the background) at 15 K and 39 K for BSTS1|Ni$_{81}$Fe$_{19}$. The voltage signal is proportional to Δ*P* at 15 K, while it exhibits a nonlinear dependence at 39 K.

Figure 4

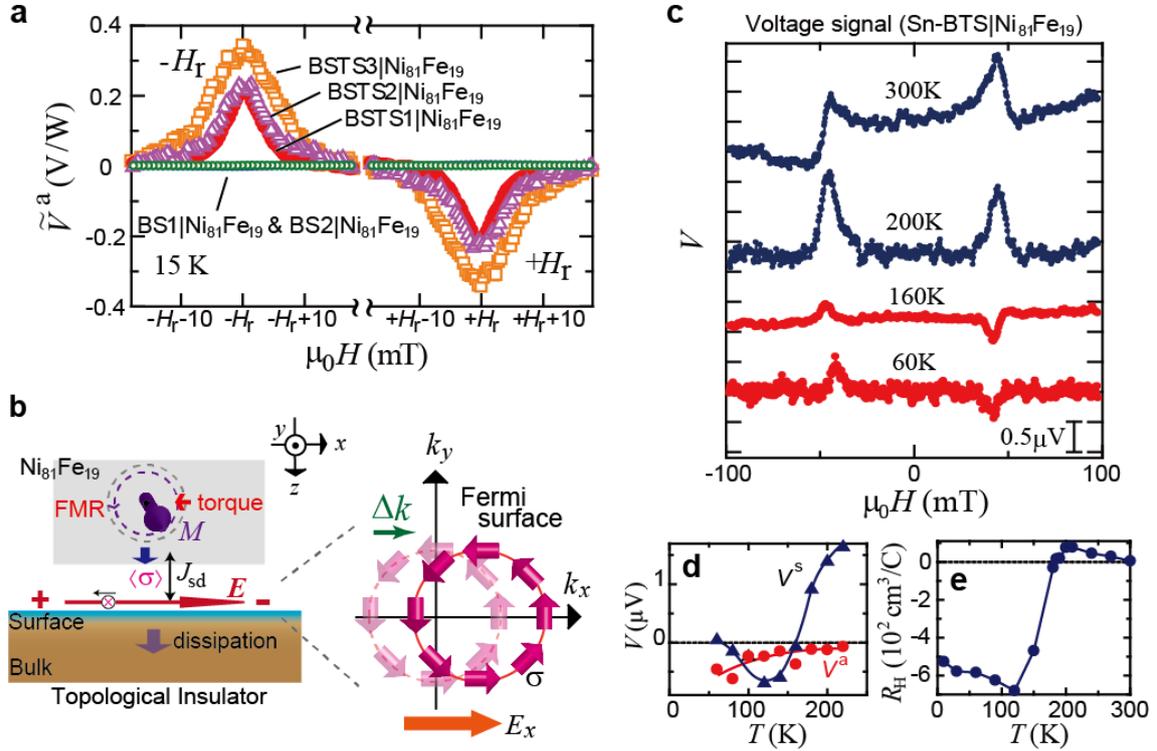

**Figure 4: Magnitude of spin-electricity conversion effect for BSTS|Ni$_{81}$Fe$_{19}$, its theoretical model, and its observation at high temperatures in Sn-BTS|Ni$_{81}$Fe$_{19}$**

(**a**) The antisymmetric part of $V$ divided by the absorption power ($\Delta P$), denoted $\tilde{V}^a$, for BS|Ni$_{81}$Fe$_{19}$ and BSTS|Ni$_{81}$Fe$_{19}$ samples at 15 K. A sizable $\tilde{V}^a$ is observed for BSTS|Ni$_{81}$Fe$_{19}$ samples, but not for BS|Ni$_{81}$Fe$_{19}$ samples. The sign of $\tilde{V}^a$ is the same for all the three BSTS samples; it is positive in $-H_r$ and negative in $+H_r$. (**b**) A schematic illustration of the theoretical mechanism for the observed effect. The exchange coupling at the interface exerts torque on the precessing Ni$_{81}$Fe$_{19}$-spins, which increases the line width of the FMR spectrum. The spin polarization induced in the topological surface state gives rise to a shift of the Fermi circle of the helical Dirac fermions toward the $x$ direction owing to the spin-momentum locking of the surface state, resulting in a charge flow. The sign of the electric field dictated by the known helicity of the surface state is consistent with the experimental results shown in (**a**). (**c**) Magnetic-field ($H$) dependence of electromotive force ($V$) at various temperatures for Sn-BTS|Ni$_{81}$Fe$_{19}$. The microwave power ($P_{in}$) was 0.05 mW. Clear antisymmetric signals were observed at 160 K and 60K. (**d**) Temperature ($T$) dependences of symmetric ($V^s$) and antisymmetric ($V^a$)

parts of $V$ for Sn-BTS|Ni$_{81}$Fe$_{19}$ below 220 K. $V^s$ shows a sign change at ~160 K, while $V^a$ is always negative. Solid curves are guides to the eyes. (**e**) $T$ dependence of the Hall coefficient ($R_H$) in Sn-BTS. $R_H$ is defined by the slope of the Hall resistivity at 5 T. $R_H$ shows a sign change at ~180 K, which signifies a crossover from $p$-type bulk-dominated transport to $n$-type surface-dominated transport.

**Supplementary Information for "Bulk topological insulators as inborn spintronics detectors"**

Y. Shiomi, K. Nomura, Y. Kajiwara, K. Eto, M. Novak, Kouji Segawa, Yoichi Ando, and E. Saitoh

**Table of contents**

---------------------------

A. Theoretical analysis of the increase in $\alpha$

The increase in the effective Gilbert damping constant $\alpha$ in the presence of BSTS can be related to a magnetic exchange coupling at the interface ($J_{sd}\vec{S}\cdot\vec{\sigma}$) [10,16,17,25]. The exchange coupling induces the torque on the Py magnetization $\vec{M}=(M/S)\vec{S}$, i.e.

$$\vec{T}_{surface} = \gamma \vec{M} \times J_{eff} \langle \vec{\sigma} \rangle \ , \tag{1}$$

where $J_{eff} = \mu_0 J_{sd} n_{Py} S / M$ is the effective exchange coupling constant [16,17] ($n_{Py}$: the number of Py atoms per unit volume). The Landau-Lifshitz-Gilbert (LLG) equation including the torque term ($\vec{T}_{surface}$) for the Py magnetization is given by

$$\frac{d\vec{M}}{dt} = \gamma \vec{M} \times \vec{B}_{eff} + \frac{\alpha_0}{M} \vec{M} \times \frac{d\vec{M}}{dt} + \vec{T}_{surface} \ . \tag{2}$$

Here, let us derive the expression of $\vec{T}_{surface}$. The Hamiltonian for the surface state is modified in the presence of the exchange coupling [S1,S2,25]:

$$H = v_F (\hat{z} \times \vec{\sigma}) \cdot \vec{p} + J_{sd} \vec{S} \cdot \vec{\sigma} = v_F (\hat{z} \times \vec{\sigma}) \cdot (\vec{p} + e\vec{a}) + J_{sd} S_z \sigma_z \ , \tag{3}$$

where $\vec{a} = (J_{sd}/ev_F)(\hat{z} \times \vec{S})$ is the effective vector potential. By comparing the following two relations about the electric current density

$$\vec{j} = -ne\left\langle \frac{\partial H}{\partial \vec{p}} \right\rangle = -nev_F(\hat{z} \times \langle \vec{\sigma} \rangle) \quad \text{and} \tag{4}$$

$$\vec{j} = \sigma_{xx}(-\dot{\vec{a}}) = \sigma_{xx}\left(-\frac{J_{sd}}{ev_F}\hat{z} \times \dot{\vec{S}}\right), \tag{5}$$

we obtain

$$\langle \vec{\sigma} \rangle = \frac{\mu_H J_{sd}}{ev_F^2}\dot{\vec{S}}. \tag{6}$$

Here, $\mu_H$ and $v_F$ are the mobility and the Fermi velocity for the surface electrons, respectively. The torque term is thereby expressed as

$$\vec{T}_{surface} = \frac{\mu_H J_{sd}^2 S}{\hbar e v_F^2 M}\vec{M} \times \dot{\vec{M}}. \tag{7}$$

The LLG equation reads

$$\frac{d\vec{M}}{dt} = \gamma \vec{M} \times \vec{B}_{eff} + \frac{\alpha_0 + \delta\alpha}{M}\vec{M} \times \frac{d\vec{M}}{dt}, \tag{8}$$

$$\delta\alpha = \frac{\mu_H J_{sd}^2 S}{\hbar e v_F^2}. \tag{9}$$

The intrinsic part of the line width ($\Delta H$) is related with $\alpha$ by $\alpha = (\gamma/2\omega)\mu_0\Delta H$. The increase in $\alpha$ ($\delta\alpha$) by spin injection into the attached BSTS is obtained as [11,S3]

$$\delta\alpha = (\gamma/2\omega)(\mu_0\Delta H_{BSTS|Py} - \mu_0\Delta H_{Py}). \tag{10}$$

From the experimental values $\mu_0\Delta H_{BSTS|Py} = 6.50$ mT and $\mu_0\Delta H_{Py} = 4.25$ mT at 5 GHz (Fig. 2**b**), $\delta\alpha$ is estimated to be $\sim 7.8 \times 10^{-3}$, which is comparable to that for Py/Pt bilayers (~0.01 [11, S3]). Using $v_F = 4.6 \times 10^5$ m/s [5], $\mu_H = 0.10$ m$^2$/Vs [5], and $S = M/(n_{Py}\mu_0\gamma\hbar) \approx 0.306$, we estimate the magnitude of $J_{sd}$ to be $|J_{sd}| = 5.9$ meV. This value is smaller than that in Mn-doped topological insulators ($|J_{sd}| \approx 100$ meV

**B**. Theoretical analysis of the spin-electricity conversion effect

When spins are polarized on the surface state, the helical Dirac dispersion results in an electric field along the Hall direction. In this section, we derive the relation between the generated electric field and the pumped spin.

The equation of motion for the surface electrons is

$$\hbar \dot{\vec{k}} = -e\vec{E} + \frac{\hbar \vec{k}}{\tau} \,, \tag{11}$$

where $\tau$ is scattering time. The shift of $k_x$ in the presence of $E_x$ is hence given by $\Delta k_x = (e\tau/\hbar)E_x$. The spin polarization $<\sigma_y>$ per unit area ($<\sigma_y>/A$) is calculated as

$$\frac{\langle \sigma_y \rangle}{A} = \frac{1}{A}\sum_{\vec{k}} f_{\vec{k}} \langle \vec{k},+|\hat{\sigma}_y|\vec{k},+\rangle = \int \frac{dk_x dk_y}{(2\pi)^2} \Delta k_x \frac{\partial f_0}{\partial k_x}\left(\frac{\vec{k}}{k}\times \hat{z}\right)_y$$
$$= -\frac{e\tau E_x}{4\pi\hbar}\int_0^\infty dk \frac{\partial f_0}{\partial k} k \approx -\frac{e\tau k_F}{4\pi\hbar}E_x \tag{12}$$

Thus, the spin polarization along $y$ axis induces the electric field along $x$ axis [26]:

$$E_x = -\frac{4\pi\hbar}{e\tau k_F}\frac{\langle \sigma_y \rangle}{A}\,. \tag{13}$$

The pumped spin rate $\Sigma$ at FMR and the spin polarization on the surface state $<\sigma_y>$ are related with each other by [1]

$$\langle \sigma_y \rangle = \eta \Sigma \tau \,, \tag{14}$$

where $\eta$ is the spin-injection efficiency (dimensionless phenomenological parameter). Equation (14) means that the spin polarization induced in the time interval $\tau$ (during which the electron is accelerated and shifts the Fermi surface) is given by the pumping rate and the spin-injection efficiency. $\eta$ may depend on the leakage of spins into the bulk electrons, on the reflection of spins to the ferromagnet, and on the effective spin-exchange coupling at the interface. From Eqs. (13) and (14), the generated transverse electric field is written as

$$E_x = -\frac{4\pi}{ek_F}\frac{\hbar}{A}\eta\Sigma \,. \tag{15}$$

The formula for $\Sigma$ has been derived and is given by [11,S5]

$$\Sigma = \frac{2}{g\mu_B} \frac{\delta\alpha}{\alpha_0 + \delta\alpha} \frac{1}{\sqrt{M^2 + 4(\omega/\gamma)^2}} \Delta P , \tag{16}$$

where $g$ is the electron g-factor. Thus, the electric field is expressed as

$$E_x = -\eta \frac{8\pi}{ek_F} \frac{\hbar}{g\mu_B} \frac{\delta\alpha}{\alpha + \delta\alpha} \frac{1}{\sqrt{M^2 + 4(\omega/\gamma)^2}} \frac{\Delta P}{A} . \tag{17}$$

The sign of the electric field is consistent with the experimental results (Figs. **2c**, **3a-3c**, **4a**, and **4c**). The experimental values are $\tilde{V}^a = E_x l / \Delta P \approx -0.2$ V/W, $\delta\alpha/(\alpha_0 + \delta\alpha) = 0.346$, $\omega = 2\pi f = 3.14 \times 10^{10}$ Hz, and $A = 1.0 \times 10^{-6}$ m$^2$. Also using $k_F = 3.9 \times 10^8$ m$^{-1}$ [5] and $g = 2.12$, we obtain $\eta = 2.3 \times 10^{-4}$.

**C**. Comments on other possible contributions to electromotive force

A charge pumping effect related with parity anomaly has been theoretically predicted in topological-insulator/ferromagnet junction structures [S6,S7]. In order to observe this effect, however, it is necessary to tune the Fermi level close to the Dirac point.

Recent reports [19,S8] pointed out possible additional conduction of topologically trivial two-dimensional electron gas (2DEG) near the surfaces of topological insulators. This 2DEG emerges due to bulk-band bending near the surfaces of topological insulators. Since this topologically-trivial surface state shows Rashba-type spin splitting [S8], this could contribute to the spin-electricity conversion effect, as mentioned in the main text. The generated electric field is, however, much smaller than that on the topological surface state with the perfect spin-momentum locking.

**D**. Fitting of FMR spectra

The FMR spectra were fitted using

$$|\Delta S_{21}(H)| = A \frac{(\Delta H/2)^2}{(H-H_r)^2 + (\Delta H/2)^2} + B \frac{-\Delta H(H-H_r)}{(H-H_r)^2 + (\Delta H/2)^2} + CH + D . \tag{18}$$

Here, $A$, $B$, $C$, $D$, and $\Delta H$ are fitting parameters. A background term is here assumed to be linear with $H$ around $H_r$ ($-H_r$). In Fig. **S1**, we show an example of FMR spectra and the fitted curves. As described in the main text, $\Delta H$ consists of the extrinsic part due to the inhomogeneity ($\Delta H_0$) and the intrinsic part ($\Delta H_i$) proportional to frequency. Frequency dependence of the intrinsic part is shown in Fig. **2b**. The values of $\mu_0 \Delta H_0$ at 15 K are 0.18 mT, 1.0 mT, and 6.6 mT for BSTS1|Py, BSTS2|Py, and BSTS3|Py,

respectively.

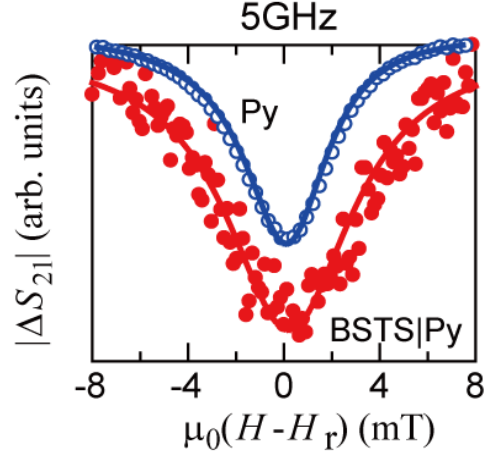

Figure **S1**: **Fitting examples for FMR spectra**

Magnetic-field ($H$) dependences of FMR spectra for Py (open circles) and BSTS1|Py (closed circles). The solid curves are the fittings (see text).

**E**. Estimation of absorption power

The microwave absorption power at FMR ($\Delta P$) was estimated using the relation of $P_{out} = |S_{21}|^2 P_{in}$, where $P_{in}$ and $P_{out}$ are incoming and outgoing power on the load, respectively [S9]. Using the resonant amplitude of $|S_{21}(H_r)|$ (Fig. **S2**), we approximately obtained $\Delta P$ as

$$\Delta P = \left( |S_{21}^{0}(H_r)|^2 - |S_{21}^{0}(H_r) - \Delta S_{21}(H_r)|^2 \right) P_{in} \approx 2|S_{21}^{0}(H_r)||\Delta S_{21}(H_r)|P_{in}. \quad (19)$$

We used output power of a vector network analyzer as the $P_{in}$ value. The value of $|S_{21}^{0}(H_r)|$ shows little change with temperature (< 100 K) or with microwave power, while it shows a large change with microwave frequency.

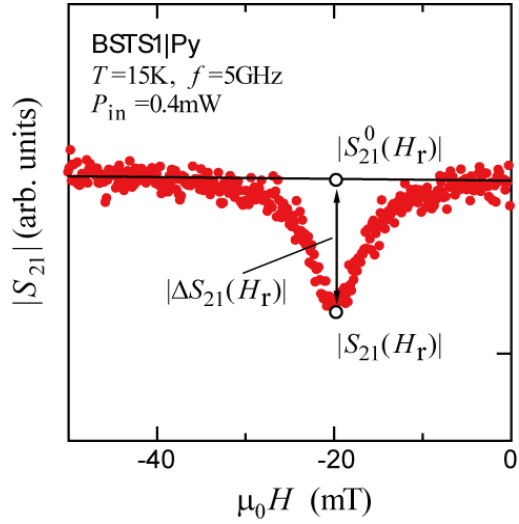

Figure **S2**: **Estimation of resonance absorption power**

An example of the magnetic-field ($H$) dependence of $|S_{21}|$ for BSTS1|Py (closed circles). The resonance absorption power ($\Delta P$) is estimated using the values of $|S_{21}|$ and the background $|S_{21}^0|$ at FMR (see text).

**F**. Power dependence of electromotive force at 15 K

The electromotive force at FMR for all the BSTS|Py samples is proportional to the absorption power $\Delta P$, as shown in Fig. **S3**.

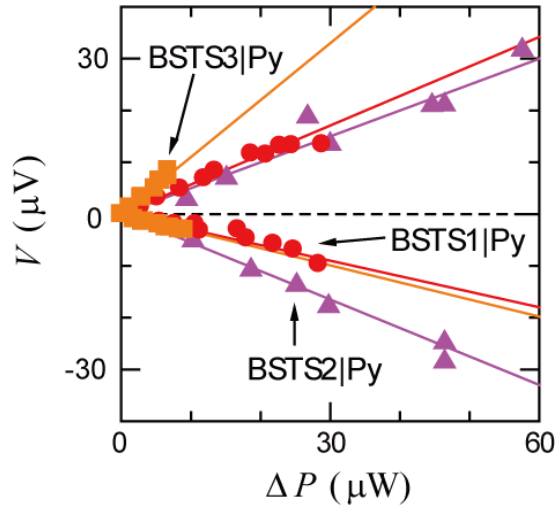

Figure **S3**: **Absorption-power dependences of electromotive force at 15 K**

Electromotive force ($V$) as a function of absorption power ($\Delta P$) for BSTS samples (BSTS1: circles, BSTS2: triangles, and BSTS3: squares) at 15 K. The solid lines are merely guides to the eyes.

G. Analysis of FMR using the Kittel formula at 15 K

The values of $M$ (the magnetization of Py) and $\gamma$ (the gyromagnetic ratio) at 15 K are obtained from the Kittel formula, *i.e.*

$$\left(\frac{\omega}{\gamma}\right)^2 = \mu_0 H_r (\mu_0 H_r + M). \qquad (20)$$

Fitting the experimental data (the solid curve in Fig. **S4**) results in $M=0.80$ T and $\gamma = 2.2 \times 10^{11}$ T$^{-1}$s$^{-1}$.

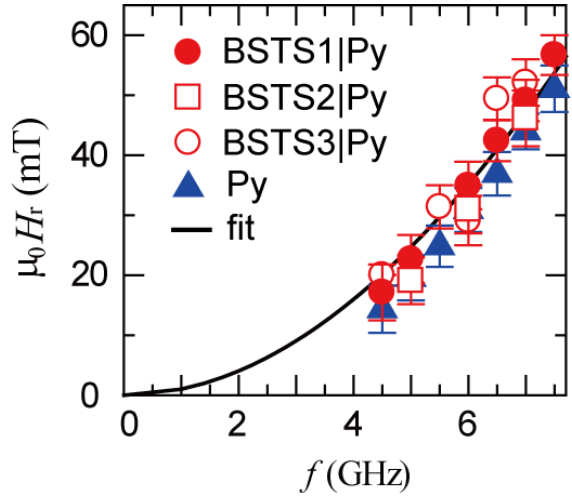

Figure **S4**: **Frequency dependence of FMR magnetic field at 15 K**

Frequency ($f$) dependence of the resonance magnetic field ($H_r$) for the BSTS|Py samples and a single Py film. The symbols are the same as those in Fig. **2b**. The solid curve is the fit to the Kittel formula.

---------------------------